\documentclass[]{article}
\usepackage{graphicx,amsmath,amssymb,amsbsy,latexsym}
\usepackage[mathscr]{eucal}

\def\tr{\mathrm{tr}}

\newcommand{\eqa}{\begin{eqnarray}}
\newcommand{\neqa}{\end{eqnarray}}
\newcommand{\be}{\begin{equation}}
\newcommand{\ee}{\end{equation}}

\newcommand{\Ref}[1]{(\ref{#1})}

\newcommand{\dual}{\,\,{}^\star\!}

\newcommand{\Hil}{\mathcal{H}}

\begin{document}

\title{Lorentzian LQG vertex amplitude}
\author{Roberto Pereira
\\[1mm]
\normalsize \em CPT%
\footnote{Unit\'e mixte de recherche (UMR 6207) du CNRS et des Universit\'es
de Provence (Aix-Marseille I), de la Meditarran\'ee (Aix-Marseille II) et du Sud (Toulon-Var); laboratoire affili\'e \`a la FRUMAM (FR 2291).} , CNRS Case 907, Universit\'e de la M\'editerran\'ee, F-13288 Marseille, EU}
\date{\small\today}

\maketitle\vspace{-7mm}

\begin{abstract}
\noindent We generalize a model recently proposed for Euclidean quantum
gravity to the case of Lorentzian signature. The main features of the Euclidean model are preserved in
the Lorentzian one. In particular, the boundary Hilbert space matches
the one of $SU(2)$ loop quantum gravity. As in the Euclidean
case, the model can be obtained from the Lorentzian Barrett-Crane model from a flipping of the Poisson structure, or alternatively, by adding a topological term to the action and taking the small Barbero-Immirzi parameter limit.
\end{abstract}

\section*{Introduction}

Loop quantum gravity (LQG) \cite{lqg} provides a well defined,
background independent, construction of the kinematical Hilbert
space of quantum general relativity. Spin foam techniques \cite{SF} have been
developed as a possible framework to study the quantum dynamics. A spin foam is a two complex (union of edges, faces and
vertices) colored by quantum numbers (faces are labelled by
representations of a given group and edges by intertwiners). It
can be interpreted as the history of a spin network (more
precisely, the boundary of a spin foam is a spin network). A spin
foam model is given by the assignment of amplitudes to faces,
edges and vertices.

The most studied model so far is the Barrett-Crane (BC) model for both Lorentzian \cite{bc_lorentz} and Euclidean \cite{bc_euclidean} signatures. It is obtained as a modification of a topological BF
quantum field theory by imposing the discrete analogues of the
constraints - called simplicity constraints -  that, in the
continuum limit, reduce BF theory to general relativity
\cite{plebanski}. Much work has been carried out in recent years
to extract the low energy behavior of this model \cite{propagator}
and it turns out that some components of the two-point functions
are in disagreement with the expected behavior determined by
standard perturbative quantum gravity \cite{emanuelle}. As argued
in \cite{eprlett,eprpap} the problem can be traced back to the way
some of the constraints are imposed in the Barrett-Crane model. In
fact, the simplicity constraints form a second class system
\cite{dirac} and in the BC model these are imposed as strong
operator equations \cite{baezbarrett}, killing then physical
degrees of freedom. In \cite{eprlett,eprpap} a reformulation of
these constraints has been proposed and this allows for a new
sector of solutions. This can be obtained from the BC model from a
flipping of the Poisson structure, or alternatively, by adding a
topological term to the action and taking the small
Barbero-Immirzi parameter limit.

In \cite{eprlett,eprpap} only the Euclidean signature case was
considered. Here we extend the construction to the Lorentzian
case. The main features of the Euclidean model are preserved in
the Lorentzian case. In particular, the boundary Hilbert space matches
the one of $SU(2)$ loop quantum gravity.

The letter is organized as follows. In the first section we review the discrete classical
theory and constraints; in the second section we give a brief
introduction to the representation theory of the Lorentz group; in the third section we quantize the theory and construct the vertex amplitude. We stress that the theory is discrete and we work with
a fixed triangulation. Matters of triangulation independence could be adressed by a group field theory
\cite{gft} approach.

\section*{Classical theory and simplicity constraints}

Following \cite{eprpap}, we introduce a discrete theory that approximates, in the continuum limit, general relativity as a constrained BF theory. The discrete theory is constructed from a Regge-like discretization of space-time. The classical discrete action that approximates BF with a topological term is given by:
\begin{eqnarray}
\nonumber
S_{disc.} &=& \frac{1}{2} \sum_{f \in int\Delta} \tr\left[B_f(t)U_f(t)
+ \frac{1}{\gamma}\dual B_f(t) U_f(t) \right]  \\
&& +\frac{1}{2} \sum_{f \in \partial\Delta} \tr\left[B_f(t)U_f(t,t')
+ \frac{1}{\gamma}\dual B_f(t) U_f(t,t') \right]
\end{eqnarray}

where $U_f(t),U_f(t,t')\in SL(2,\mathbb{C})$, $B_f(t)\in \mathfrak{sl}(2,\mathbb{C})$, and $\gamma\in\mathbb{R}_{*}^+$ is the Barbero-Immirzi parameter. For the definition of these variables and details on the construction, see \cite{eprpap}.

The boundary phase space is parameterized by the pairs
$\left(B_f(t),U_f(t,t')\right)$, one for each boundary link.
Because of the introduction of the topological term, the variable
conjugate to $U_f(t,t')$ is
\begin{equation}
J_f(t) = B_f(t) + \frac{1}{\gamma}\dual B_f(t)
\end{equation}
where $\dual\;$ stands for the Hodge dual in the Lorentz internal algebra. More precisely, the matrix elements $J_f(t)^{IJ}$ ($I,J=0,..3$ are the indices of the internal Lorentz algebra) have as their Hamiltonian vector fields the right invariant vector fields
on the group $U_f(t,t')$. Inverting this equation gives
\begin{equation}
B_f(t) = \left(\frac{\gamma^2}{\gamma^2+1}\right)
\left(J_f(t)-\frac{1}{\gamma}\dual J_f(t)\right). \label{Bf}
\end{equation}
For the cases $\gamma \ll 1$ and $\gamma \gg 1$,
this reduces to
\begin{align}
\nonumber \gamma \ll 1 &\;\,\,\leadsto\;\,\,
B = -\gamma \dual J &\qquad \gamma \gg 1
&\;\,\,\leadsto\;\,\, B = J .
\end{align}
corresponding respectively to the flipped and non flipped Poisson
structures of $SL(2,\mathbb{C})$.

Next, we impose the discrete analog of the constraints that
reduce BF theory to GR. The analog of the Gauss law in the continuum theory is given by the closure constraint (one per tetrahedron $t$):
\be
\sum_{f\in t}\;J_f(t)=0.
\ee
As argued in \cite{eprpap}, these will be imposed automatically by the dynamics. In addition, we impose the simplicity constraints. They can be cast into the form (\cite{eterabf} and \cite{area_lett}):
\begin{eqnarray}
C_{ff}&:=&\dual J_f\cdot J_f\left(1-\frac{1}{\gamma^2}\right)+\frac{2}{\gamma}J_f\cdot J_f \approx 0 \label{C}\\
C_f^J&:=&n_I\left((\dual J_f)^{IJ}+\frac{1}{\gamma}J_f^{IJ}\right) \approx 0 \label{CJ}
\end{eqnarray}
where in the second equation the vector $n_I$ is the same for all
faces meeting in a given tetrahedron of $\Delta$. These will be imposed directly on the quantum theory.

In order to proceed, let us fix $n_I=\delta_I^0$. The general case
will be recovered by gauge invariance. Remark that with this
choice of $n_I$ we are also constraining the tetrahedra to be
space-like. Equation \Ref{CJ} then becomes

\begin{equation}
C_f^j=\frac{1}{2}\epsilon^j{}_{kl}J_f^{kl}+\frac{1}{\gamma}J^{0j}=L_f^j+\frac{1}{\gamma}K_f^j \approx 0 \label{Cj}
\end{equation}

where $\epsilon^{j}{}_{kl}:=\epsilon^{0j}{}_{kl}$,
$L_f^j:=\frac{1}{2}\epsilon^{j}{}_{kl}J_f^{kl}$, $K_f^j:=J^{0j}$.
$L_f^i$ and $K_f^i$ are resp. the usual rotation and boost
generators of the Lorentz algebra. We take (\ref{C},\ref{Cj}) as
our basic set of constraints. The Poisson algebra of the
constraints can be carried out easily and it closes only in the
large $\gamma$  limit (see \cite{elpr} for a discussion for
general values of $\gamma$).

\section*{Representation theory of the Lorentz group}

In this section, we review some facts on the Lorentzian
representation theory (for details see \cite{ruhl}). There are two
useful representation spaces. The first is the space of
homogeneous functions on two complex variables:

\be
  f(\lambda z_1,\lambda z_2)=\lambda^a \bar{\lambda}^b f(z_1,z_2)
\ee

where $(a,b)$ is the degree and $(a-b) \in \mathbb{Z}$.
Irreducible unitary representations of the principal series (which
is the one that appears in the decomposition of the regular
representation) are given by $f$ homogeneous of degree $((i\rho-n)/2-1,(i\rho+n)/2-1)$, $n\in
\mathbb{N}$ and $\rho \in \mathbb{R}$. The group is represented in this space as:

\be
T_{n\rho}(g)f(z_1,z_2)=f(\alpha z_1+\gamma z_2,\beta z_1+\delta z_2)
\ee

where

\be
g \in SL(2,\mathbb{C})\;=\;\left(\begin{array}{cc} \alpha & \beta \\ \gamma & \delta\end{array}\right).
\ee

A bi-invariant, invariant under inversion measure can also be
written. Square integrable functions have a well defined Fourier
transform and a Plancherel theorem can be written, but it will be
more useful to write them down in the second representation space
mentioned above. This is given by restricting $(z_1, z_2)$ to the
sphere $|z_1|^2+|z_2|^2=1$. The restriction is possible because of
the homogeneity of $f$. We then write $u\in SU(2)$:

\be u=\left(\begin{array}{cc} \bar{z}_2 & -\bar{z}_1 \\ z_1 & z_2 \end{array}\right).
\ee

and the functions $\phi(u):=f(u_{21},u_{22})$ generate the
representation space. The Peter-Weyl theorem gives the
decomposition of $\phi(u)$ into the $SU(2)$ representation matrices:

\be
\phi(u)=\sum_{j\geq n/2}\sum_{|q|\leq j}\;d^j_q\;\phi^j_q(u)
\ee
where $\phi^j_q(u):=\sqrt{2j+1}D^j_{{\scriptstyle \frac{n}{2}}q}(u)$. This gives an explicit formula for the decomposition of the representation space $\mathcal{H}_{(n,\rho)}$ into $SU(2)$ irreducibles when viewed as a reducible representation space under the action of a $SU(2)$ subgroup: $\mathcal{H}_{(n,\rho)}=\bigoplus_{j\geq n/2}\;\mathcal{H}_j$. The basis $\left\{\phi^j_q\right\}$ is referred to as the canonical basis in the literature. This is the basis that diagonalizes simultaneously the operators $\left\{J\cdot J,\dual J\cdot J,L^2,L^z\right\}$.

We complete this section with some useful formulas. First the
character decomposition of the delta function on the group:

\be
\delta(g)=\sum_{n=0}^{\infty}\int_{-\infty}^{\infty}(n^2+\rho^2)d\rho\;\left[\sum_{j\geq n/2,|q|\leq j}\bar{D}^{n,\rho}_{jqjq}(g)\right] \label{resid}
\ee

where

\be
D^{n\rho}_{jqj'q'}(g):=\int_{SU(2)}du\;\bar{\phi}^j_q(u)\; T_{n\rho}(g)\cdot\phi^{j'}_{q'}(u).
\ee

The restriction of the $D$ matrices to the $SU(2)$ subgroup is also important:

\be
D^{n\rho}_{jqj'q'}(u)=\delta_{jj'}D^j_{qq'}(u) \label{restD}
\ee

where the second $D$ matrix is the usual $SU(2)$ representation matrix.

Next, the integration over four group elements gives a sum over intertwiners\footnote{Note that, because of the non-compactedness of the group, any averaging procedure has to be understood as formal. In special the invariant tensors used here have to be understood as generalized tensors of infinite norm. Some regularization procedure is expected at the end to make the amplitudes finite.}:

\be
\int_{SL(2,\mathbb{C})}dg\; D^{n_1\rho_1}_{j_1q_1j'_1q'_1}(g)\;D^{n_2\rho_2}_{j_2q_2j'_2q'_2}(g)\;\bar{D}^{n_3\rho_3}_{j_3q_3j'_3q'_3}(g)\;\bar{D}^{n_4\rho_4}_{j_4q_4j'_4q'_4}(g)=\sum_n\int d\rho(n^2+\rho^2)\;C^{n\rho}_{(j_1q_1)...(j_4q_4)}\bar{C}^{n\rho}_{(j'_1q'_1)...(j'_4q'_4)} \label{intert}
\ee

Here $C^{n\rho}_{(j_1q_1)...(j_4q_4)}$ is the intertwiner labelled
by $(n,\rho)$ between representations
$(n_1,\rho_1)...(n_4,\rho_4)$. It is defined as:

\be
C^{n\rho}_{(j_1q_1)...(j_4,q_4)}=\sum_{j,q}\; C^{n_1\rho_1 n_2\rho_2 n\rho}_{(j_1q_1) (j_2q_2) (jq)}\bar{C}^{n_3\rho_3 n_4\rho_4 n\rho}_{(j_3q_3) (j_4q_4) (jq)}
\ee

and $C^{n_1\rho_1 n_2\rho_2 n\rho}_{(j_1q_1) (j_2q_2) (jq)}$ are
the Clebsch-Gordan coefficients for the Lorentz group
\cite{andersonetal}. They satisfy orthogonality:

\be
\sum_{j_1q_1j_2q_2}\;C^{n_1\rho_1 n_2\rho_2 n\rho}_{(j_1q_1) (j_2q_2) (jq)}\bar{C}^{n_1\rho_1 n_2\rho_2 n'\rho'}_{(j_1q_1) (j_2q_2) (j'q')}=\frac{\delta (\rho-\rho')}{n^2+\rho^2}\;\delta_{nn'}\delta_{jj'}\delta_{qq'};
\ee

and completeness:

\be
\sum_n\int d\rho (n^2+\rho^2)\sum_{jq}\; C^{n_1\rho_1 n_2\rho_2 n\rho}_{(j_1q_1) (j_2q_2)(jq)}\bar{C}^{n_1\rho_1 n_2\rho_2 n\rho}_{(j'_1q'_1) (j'_2q'_2) (jq)}=\delta_{j_1j'_1}\delta_{j_2j'_2}\delta_{q_1q'_1}\delta_{q_2q'_2}.
\ee

This last equation and the intertwining property of the
Clebsch-Gordan coefficients are sufficient to prove the following
identity:

\be
D^{n_1\rho_1}_{j_1q_1j'_1q'_1}(g)D^{n_2\rho_2}_{j_2q_2j'_2q'_2}(g)=\sum_n\int d\rho (n^2+\rho^2)\sum_{jq}C^{n_1\rho_1 n_2\rho_2 n\rho}_{(j_1q_1) (j_2q_2) (jq)}D^{n\rho}_{jqj'q'}(g)\bar{C}^{n_1\rho_1 n_2\rho_2 n\rho}_{(j'_1q'_1) (j'_2q'_2) (j'q')}\ee

which can then be used to prove \Ref{intert}. Normalization of $D$ matrices is such that:

\be
\int_{SL(2,\mathbb{C})} dg\; \bar{D}^{n_1\rho_1}_{j_1q_1j'_1q'_1}(g)\; D^{n_2\rho_2}_{j_2q_2j'_2q'_2}(g)=\frac{\delta(\rho-\rho')}{n^2+\rho^2}\;\delta_{nn'}\delta_{(j_1q_1)(j_2q_2)}\delta_{(j'_1q'_1)(j'_2q'_2)}.
\ee

Finally, the Casimir operators for a representation in the principal series $(n,\rho)$ are given by:
\begin{eqnarray}
C_1&=&J\cdot J=2\left(L^2-K^2\right)=\frac{1}{2}\left(n^2-\rho^2-4\right); \label{C1} \\
C_2&=&\dual J\cdot J=-4L\cdot K=n\rho. \label{C2}
\end{eqnarray}

\section*{Quantization and vertex amplitude}

From the discrete boundary variables and their symplectic
structure, one can write the Hilbert space associated
with a boundary or 3-slice.
To do this, it is simpler to
switch to the dual, 2-complex picture, $\Delta^*$.
For each 3-surface $\Sigma$ intersecting no vertices of $\Delta^*$,
let $\gamma_{\Sigma} := \Sigma \cap \Delta^*$.  The Hilbert space
associated with $\Sigma$ is then
\begin{equation}
\Hil_{\Sigma} = L^2\left(SL(2,\mathbb{C})^{\times L}\right)
\end{equation}
where $L$ is the number of links in the triangulation $\gamma_\Sigma$. Let $\hat{J}_f(t)^{IJ}$ denote the
right-invariant vector fields, determined by the
basis $J^{IJ}$ of $\mathfrak{sl}(2,\mathbb{C})$, on the copy of
$SL(2,\mathbb{C})$ associated with the link $l = f \cap \Sigma$
determined by $f$, with orientation such that the node
$n = t \cap \Sigma$ is the source of $l$.
From eq. \Ref{Bf}, the $B_f(t)$'s are then quantized as
\begin{equation}
\hat{B}_f(t) := \left(\frac{\gamma^2}{\gamma^2+1}\right)
\left(\hat{J}_f(t)-\frac{1}{\gamma}\dual \hat{J}_f(t)\right).
\end{equation}

Next we promote \Ref{C} and \Ref{Cj} to quantum operators. We note
that the first constraint commutes with the others and can be
carried directly to quantum theory. For either $\gamma \ll 1$ or
$\gamma \gg 1$ this condition is satisfied by the simple
representations of $SL(2,\mathbb{C})$, i.e., for $n\rho=0$, which
has two distinct classes of solutions given by the representations
labelled by either $(n,0)$ or $(0,\rho)$.

For large $\gamma$, the constraint algebra closes and the
off-diagonal simplicity constraints can be imposed strongly as
operator equations. The solution is given by restricting to the
$SU(2)$ invariant subspace, which appears only in the
decomposition of the representations $(0,\rho)$. This is the
Lorentzian Barrett-Crane model \cite{bc_lorentz} (see also
\cite{etera_projsn}).

For small $\gamma$, the algebra does not close and we need to
impose the constraints in a weaker sense. We follow the strategy
in \cite{eprlett,eprpap} and impose $M_f:=(C_f^i)^2\approx 0$,
allowing at the same time for possible corrections, small in the
semiclassical limit. In this sector the constraint reads: \be
M_f=(K_f^i)^2\approx 0. \label{master} \ee Using eq. \Ref{C1} we
see that, up to semiclassical corrections, the solution is given
by choosing the simple representations of the form $(n,0)$ and
restricting to the lowest $SU(2)$ irreducible in its
decomposition, that is, $j=n/2$. This defines the projection from
the $SL(2,\mathbb{C})$ boundary Hilbert space to the $SU(2)$ space.
For a single $D$ matrix, this projection reads:
\begin{eqnarray}
\pi\; &:&\; L^2\left(SL(2,\mathbb{C})\right)\longrightarrow L^2\left(SU(2)\right) \nonumber \\
      &&D^{n\rho}_{jqj'q'}(g)\longmapsto D^{n/2}_{qq'}(u) \label{proj}
\end{eqnarray}
where $g\in SL(2,\mathbb{C})$ and $u\in SU(2)$ and we have used eq. \Ref{restD}. This also
defines an embedding from the $SU(2)$ Hilbert space to the $SL(2,\mathbb{C})$ space, given by inclusion followed by group
averaging over the Lorentz group. This last statement holds for
the Hilbert space associated to a single link in the boundary
triangulation. In order to make sense of it for the complete space
$\mathcal{H}_\Sigma$ we have to define the projection for
intertwiners.

Consider then four links meeting at a given node $e$
of $\gamma_\Sigma$ and labelled by simple representations
$(n_1,\rho_1)...(n_4,\rho_4)$ (where $n_i\rho_i=0$). We start with the auxiliary Hilbert space of tensors between these
representations: $\mathcal{H}_0:=\mathcal{H}_{{\scriptstyle
(n_1,\rho_1)}}\otimes ...\otimes \mathcal{H}_{{\scriptstyle
(n_4,\rho_4)}}$. Construct the constraint $C:=\sum_{i=1}^4 M_{f_i}\approx 0$. Imposing it selects, in each link, the lowest $SU(2)$ irreducible along with the simple representations
of the form $(n_i,0)$. The last step is group averaging over
$SL(2,\mathbb{C})$, which then defines the physical intertwiner
space for this node. The projection is given by:
\begin{eqnarray}
\pi\; &:&\; Inv_{SL(2,\mathbb{C})}\left(\mathcal{H}_0\right)\longrightarrow Inv_{SU(2)}\left(\mathcal{H}_{\frac{n_1}{2}}\otimes ...\otimes\mathcal{H}_{\frac{n_4}{2}}\right) \nonumber \\
&& C^{n_e\rho_e}_{(j_1,q_1)...(j_4,q_4)}\longmapsto
C^{n_e\rho_e}_{(\frac{n_1}{2},q_1),...(\frac{n_4}{2},q_4)}.
\end{eqnarray}
The embedding is given by:
\begin{eqnarray}
f\; &:&\;  Inv_{SU(2)}\left(\mathcal{H}_{\frac{n_1}{2}}\otimes
...\otimes\mathcal{H}_{\frac{n_4}{2}}\right)\longrightarrow  Inv_{SL(2,\mathbb{C})}\left(\mathcal{H}_0\right) \nonumber \\
&& i^{m_1...m_4}\longmapsto\int_{SL(2,\mathbb{C})}\;dg\;\left(\bigotimes_{i=1}^{i=4}\;D^{(2j_i,0)}(g)^{(j'_i,m'_i)}{}_{(j_i,m_i)}\right)\;
i^{m_1...m_4}.
\end{eqnarray}
Composing the embedding for
intertwiners and $D$ matrices gives an embedding of the LQG Hilbert space into the kinematical boundary space of the model. The image of this embedding is the physical Hilbert space associated to this triangulation.

We are now ready to define the vertex. We follow the same steps taken to define the Euclidean vertex \cite{eprpap}. We start with the BF amplitude for a single 4-simplex:

\be A(g_{ab})=\int_{SL(2\mathbb{C})^{\times 5}} \prod_a dV_a \;
\prod_{(ab)}\delta(V_a g_{ab} V_b^{-1}) \ee
where the indices $a,b=1,...,5$ label the tetrahedra on the boundary of the
4-simplex and $(ab)$ labels the faces between the corresponding
tetrahedra. Then use \Ref{resid} to decompose the delta function into characters and integrate to get intertwiners, using
\Ref{intert}. The amplitude can be written as:
\be
A(g_{ab})=\sum_{n_{ab}}\int d\rho_{ab}(n_{ab}^2+\rho_{ab}^2)\sum_{n_a}\int d\rho_a(n_a^2+\rho_a^2)\;15j\left((n_{ab},\rho_{ab});(n_a,\rho_a)\right)\Psi_{n,\rho}(g_{ab})
\ee
where $\Psi_{n,\rho}(g_{ab})$ denotes the spin-net functional defined as $\Psi_{n,\rho}(g_{ab}):=\bigotimes_a\;\bar{C}^{n_a\rho_a}\;\cdot\;\bigotimes_{(ab)}\; D^{n_{ab}\rho_{ab}}(g_{ab})$, where contraction follows the combinatorics of the boundary graph of a 4-simplex and indices have been omitted. Now come the simplicity constraints. These are solved by projecting as in \Ref{proj}. The final
amplitude is given by taking the $SU(2)$ scalar product of
the projected $A(g_{ab})$ with a $SU(2)$ spin-net
$\psi_{j_{ab},i_a}(g_{ab})$ on the boundary of the 4-simplex. Explicitly:

\be
A(j_{ab},i_a)=\sum_{n_a}\int d\rho_a (n_a^2+\rho_a^2)\;15j\left((2j_{ab},0);(n_a,\rho_a)\right)\; f^{i_a}_{n_a\rho_a}(j_{ab})
\ee
where
\be
f^{i_a}_{n_a\rho_a}:=i_a^{m_1...m_4}\bar{C}^{n_a\rho_a}_{(j_1,m_1)...(j_4,m_4)}
\ee
where $j_1...j_4$ are the four (fixed) $SU(2)$ representations meeting at the node $a$. The final partition function, for an arbitrary triangulation, is
given by gluing these amplitudes together with suitable edge and
face amplitudes. It can be written as:
\be
Z=\sum_{j_f,i_e} \prod_f\;(2j_f)^2\;\prod_v\;A(j_f,i_e)
\ee
where the sum is over $SU(2)$ representation labels.

\section*{Conclusion}

We have extended the construction given in \cite{eprlett,eprpap}
to the Lorentzian case. The construction given
above depends on the choice of \emph{spacelike} tetrahedra, which
is the good choice if one hopes to make contact with LQG. The
boundary Hilbert space matches the one of LQG (defined on a fixed
graph), which was our main motivation to define the model.
Relation to general relativity in an appropriate limit is still
missing and we expect that the study of the semiclassical limit of
the model as well as its $n$-point functions will shed some light
on this issue.

We close with some remarks.

\begin{itemize}
\item[(i)] First, the vertex amplitude given above as it stands is
only formal and some regularization procedure is expected in order
to make it finite, as is the case for the Lorentzian Barrett-Crane
model.

\item[(ii)] Second, the theory can actually be defined for general
values of $\gamma$ \cite{elpr}, conserving its main features.

\item[(iii)] Finally, the model defined here is clearly distinct
from other models proposed in the literature as alternatives to
the Barrett-Crane model \cite{pr_lorentz31,fk2007,sergei2007}, as
it is based solely on the simple representations of the form
$(n,0)$. Coherent states \cite{ls_cohstates,ls_model} might also
be used to give an equivalent derivation, as it is the case for
the Euclidean vertex, but we leave this to further investigation.
\end{itemize}
\vspace{0.5cm} \centerline{--------} \vspace{0.5cm}

Thanks to Alejandro Perez for suggesting the use of the discrete series, to Carlo Rovelli for a careful reading of the manuscript and to Jonathan Engle for useful discussions.

\end{document}